\def\year{2019}\relax
\documentclass[letterpaper]{article} 
\usepackage{cite}
\usepackage{aaai19}  
\usepackage{times}  
\usepackage{helvet}  
\usepackage{courier}  
\usepackage{url}  
\usepackage{graphicx}  

\usepackage{amsthm}
\usepackage{amsmath,amssymb,amsfonts}
\usepackage{booktabs}

\usepackage{textcomp}
\usepackage{xcolor}
\usepackage[font=footnotesize]{subfig}
\usepackage{multirow}
\usepackage{float}

\begin{document}

\title{Session-based Recommendation with Graph Neural Networks}

\author{
Shu Wu,\textsuperscript{\rm 1,2}
Yuyuan Tang,\textsuperscript{\rm 3}
Yanqiao Zhu,\textsuperscript{\rm 4}
Liang Wang,\textsuperscript{\rm 1,2}
Xing Xie,\textsuperscript{\rm 5}
Tieniu Tan\textsuperscript{\rm 1,2}\\
\textsuperscript{\rm 1}{Center for Research on Intelligent Perception and Computing}\\{National Laboratory of Pattern Recognition, Institute of Automation, Chinese Academy of Sciences}\\
\textsuperscript{\rm 2}{University of Chinese Academy of Sciences}\\
\textsuperscript{\rm 3}{School of Computer and Communication Engineering, University of Science and Technology Beijing}\\
\textsuperscript{\rm 4}{School of Software Engineering, Tongji University}\\
\textsuperscript{\rm 5}{Microsoft Research Asia}\\
shu.wu@nlpr.ia.ac.cn, tangyyuanr@gmail.com, sxkdz@tongji.edu.cn,\\
wangliang@nlpr.ia.ac.cn, xing.xie@microsoft.com, tnt@nlpr.ia.ac.cn
}

\maketitle

\begin{abstract}

The problem of session-based recommendation aims to predict user actions based on anonymous sessions. Previous methods model a session as a sequence and estimate user representations besides item representations to make recommendations. Though achieved promising results, they are insufficient to obtain accurate user vectors in sessions and neglect complex transitions of items. To obtain accurate item embedding and take complex transitions of items into account, we propose a novel method, i.e. {\it Session-based Recommendation with Graph Neural Networks}, SR-GNN for brevity. In the proposed method, session sequences are modeled as graph-structured data. Based on the session graph, GNN can capture complex transitions of items, which are difficult to be revealed by previous conventional sequential methods. Each session is then represented as the composition of the global preference and the current interest of that session using an attention network. Extensive experiments conducted on two real datasets show that SR-GNN evidently outperforms the state-of-the-art session-based recommendation methods consistently.

\end{abstract}

\section{Introduction}

With the rapid growth of the amount of information on the Internet, recommendation systems become fundamental for helping users alleviate the problem of information overload and select interesting information in many Web applications, e.g., search, e-commerce, and media streaming sites. Most of the existing recommendation systems assume that the user profile and past activities are constantly recorded. However, in many services, user identification may be unknown and only the user behavior history during an ongoing session is available. Thereby, it is of great importance to model limited behavior in one session and generate the recommendation accordingly. Conversely, conventional recommendation methods relying on adequate user-item interactions have problems in yielding accurate results under this circumstance.

Due to the highly practical value, increasing research interests in this problem can be observed, and many kinds of proposals for session-based recommendation have been developed. Based on Markov chains, some work \cite{Shani:2002:MRS:2073876.2073930,rendle2010factorizing} predicts the user's next behavior based on the previous one. With a strong independence assumption, independent combinations of the past components confine the prediction accuracy.

In recent years, the majority of research \cite{DBLP:journals/corr/HidasiKBT15,Tan:2016:IRN:2988450.2988452,Tuan:2017:CNS:3109859.3109900,Li:2017:NAS:3132847.3132926} apply Recurrent Neural Networks (RNNs) for session-based recommendation systems and obtain promising results. The work \cite{DBLP:journals/corr/HidasiKBT15} proposes a recurrent neural network approach at first, then the model is enhanced by data augmentation and considering temporal shift of user behavior \cite{Tan:2016:IRN:2988450.2988452}. Recently, NARM \cite{Li:2017:NAS:3132847.3132926} designs a global and local RNN recommender to capture user's sequential behavior and main purposes simultaneously. Similar to NARM, STAMP \cite{Liu:2018:SSA:3219819.3219950} also captures users' general interests and current interests, by employing simple MLP networks and an attentive net.

\begin{figure*}[h]
	\centering
	\includegraphics[width=\textwidth]{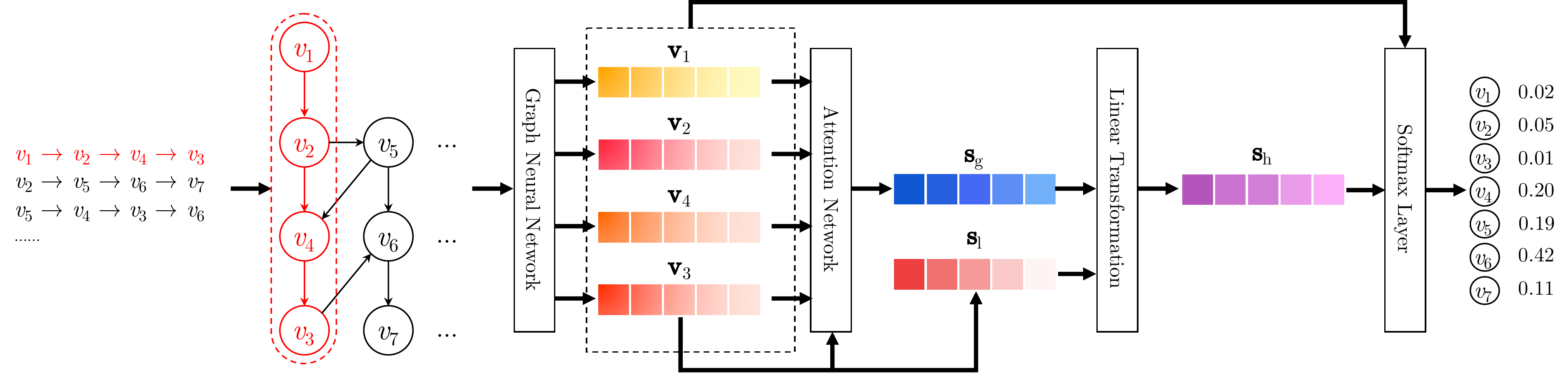}
	\caption{The workflow of the proposed SR-GNN method. We model all session sequences as session graphs. Then, each session graph is proceeded one by one and the resulting node vectors can be obtained through a gated graph neural network. After that, each session is represented as the combination of the global preference and current interests of this session using an attention net. Finally, we predict the probability of each item that will appear to be the next-click one for each session.}
	\label{fig:workflow}
\end{figure*}

Although the methods above achieve satisfactory results and become the state-of-the-arts, they still have some limitations. {\it Firstly}, without adequate user behavior in one session, these methods have difficulty in estimating user representations. Usually, the hidden vectors of these RNN methods are treated as the user representations, such that recommendations can be then generated based on these representations, for instance, the global recommender of NARM. In session-based recommendation systems, however, sessions are mostly anonymous and numerous, and user behavior implicated in session clicks is often limited. It is thus difficult to accurately estimate the representation of each user from each session. {\it Secondly}, previous work reveals that patterns of item transitions are important and can be used as a local factor \cite{Li:2017:NAS:3132847.3132926,Liu:2018:SSA:3219819.3219950} in session-based recommendation, but these methods always model single-way transitions between consecutive items and neglect the transitions among the contexts, i.e. other items in the session. Thus, complex transitions among distant items are often overlooked by these methods.

To overcome the limitations mentioned above, we propose a novel method for {\underline S}ession-based {\underline R}ecommendation with {\underline G}raph {\underline N}eural {\underline N}etworks, SR-GNN for brevity, to explore rich transitions among items and generate accurate latent vectors of items. Graph Neural Networks (GNNs) \cite{4700287,DBLP:journals/corr/LiTBZ15} are designed for generating representations for graphs. Recently, it has been employed to model graph-structured dependencies for natural language processing and computer vision applications flourishingly, e.g., script event prediction \cite{EEG2018}, situation recognition \cite{8237710}, and image classification \cite{8099493}. For the session-based recommendation, we first construct directed graphs from historical session sequences. Based on the session graph, GNN is capable of capturing transitions of items and generating accurate item embedding vectors correspondingly, which are difficult to be revealed by the conventional sequential methods, like MC-based and RNN-based methods. Based on accurate item embedding vectors, the proposed SR-GNN constructs more reliable session representations and the next-click item can be inferred.

Figure \ref{fig:workflow} illustrates the workflow of the proposed SR-GNN method. At first, all session sequences are modeled as directed session graphs, where each session sequence can be treated as a subgraph. Then, each session graph is proceeded successively and the latent vectors for all nodes involved in each graph can be obtained through gated graph neural networks. After that, we represent each session as a composition of the global preference and the current interest of the user in that session, where these global and local session embedding vectors are both composed by the latent vectors of nodes. Finally, for each session, we predict the probability of each item to be the next click. Extensive experiments conducted on real-world representative datasets demonstrate the effectiveness of the proposed method over the state-of-arts. The main contributions of this work are summarized as follows:
\begin{itemize}
	\item We model separated session sequences into graph-structured data and use graph neural networks to capture complex item transitions. To best of our knowledge, it presents a novel perspective on modeling in the session-based recommendation scenario.
	\item To generate session-based recommendations, we do not rely on user representations, but use the session embedding, which can be obtained merely based on latent vectors of items involved in each single session.
	\item Extensive experiments conducted on real-world datasets show that SR-GNN evidently outperforms the state-of-art methods.
\end{itemize}

To make our results fully reproducible, all the relevant source codes have been made public at \url{https://github.com/CRIPAC-DIG/SR-GNN}.

The rest of this paper is organized as follows. We review prior related literature in Section 2. Section 3 presents the proposed method of session-based recommendation with graph neural networks. Detailed experiment results and analysis are shown in Section 4. Finally, we conclude this paper in Section 5.

\section{Related Work}

In this section, we review some related work on session-based recommendation systems, including conventional methods, sequential methods based on Markov chains, and RNN-based methods. Then, we introduce the neural networks on graphs.

{\bf Conventional recommendation methods.}
Matrix factorization \cite{mnih2007probabilistic,koren2009matrix,koren2011advances} is a general approach to recommendation systems. The basic objective is to factorize a user-item rating matrix into two low-rank matrices, each of which represents the latent factors of users or items. It is not very suitable for the session-based recommendation, because the user preference is only provided by some positive clicks. The item-based neighborhood methods \cite{Sarwar2001} is a natural solution, in which item similarities are calculated on the co-occurrence in the same session. These methods have difficulty in considering the sequential order of items and generate prediction merely based on the last click.

Then, the sequential methods based on Markov chains are proposed, which predict users' next behavior based on the previous ones. Treating recommendation generation as a sequential optimization problem, \citeauthor{Shani:2002:MRS:2073876.2073930} (\citeyear{Shani:2002:MRS:2073876.2073930}) employ Markov decision processes (MDPs) for the solution. Via factorization of the personalized probability transition matrices of users, FPMC \cite{rendle2010factorizing} models sequential behavior between every two adjacent clicks and provides a more accurate prediction for each sequence. However, the main drawback of Markov-chain-based models is that they combine past components independently. Such an independence assumption is too strong and thus confines the prediction accuracy.

{\bf Deep-learning-based methods.}
Recently, some prediction models, especially language models \cite{mikolov2013distributed} are proposed based on neural networks. Among numerous language models, the recurrent neural network (RNN) has been the most successful one in modeling sentences \cite{mikolov2010recurrent} and has been flourishingly applied in various natural language processing tasks, such as machine translation \cite{cho2014learning}, conversation machine \cite{serban2016building}, and image caption \cite{mao2014deep}. RNN also has been applied successfully in numerous applications, such as the sequential click prediction \cite{zhang2014sequential}, location prediction \cite{liu2016strnn}, and next basket recommendation \cite{yu2016dream}.

For session-based recommendation, the work of \cite{DBLP:journals/corr/HidasiKBT15} proposes the recurrent neural network approach, and then extends to an architecture with parallel RNNs \cite{Hidasi:2016:PRN:2959100.2959167} which can model sessions based on the clicks and features of the clicked items. After that, some work is proposed based on these RNN methods. \citeauthor{Tan:2016:IRN:2988450.2988452} (\citeyear{Tan:2016:IRN:2988450.2988452}) enhances the performance of recurrent model by using proper data augmentation techniques and taking temporal shifts in user behavior into account. \citeauthor{Jannach:2017:RNN:3109859.3109872} (\citeyear{Jannach:2017:RNN:3109859.3109872}) combine the recurrent method and the neighborhood-based method together to mix the sequential patterns and co-occurrence signals. \citeauthor{Tuan:2017:CNS:3109859.3109900} (\citeyear{Tuan:2017:CNS:3109859.3109900}) incorporates session clicks with content features, such as item descriptions and item categories, to generate recommendations by using 3-dimensional convolutional neural networks. Besides, A list-wise deep neural network \cite{Wu:2017:SIE:3132847.3133163} models the limited user behavior within each session, and uses a list-wise ranking model to generate the recommendation for each session. Furthermore, a neural attentive recommendation machine with an encoder-decoder architecture, i.e. NARM \cite{Li:2017:NAS:3132847.3132926}, employs the attention mechanism on RNN to capture users' features of sequential behavior and main purposes. Then, a short-term attention priority model (STAMP) \cite{Liu:2018:SSA:3219819.3219950} using simple MLP networks and an attentive net, is proposed to efficiently capture both users' general interests and current interests.

{\bf Neural network on graphs.}
Nowadays, neural network has been employed for generating representation for graph-structured data, e.g., social network and knowledge bases. Extending the word2vec \cite{mikolov2013distributed}, an unsupervised algorithm DeepWalk \cite{Perozzi:2014:DOL:2623330.2623732} is designed to learn representations of graph nodes based on random walk. Following DeepWalk, unsupervised network embedding algorithms LINE \cite{Tang:2015:LLI:2736277.2741093} and node2vec \cite{Grover:2016:NSF:2939672.2939754} are most representative methods. On the another hand, the classical neural network CNN and RNN are also deployed on graph-structured data. \cite{Duvenaud:2015:CNG:2969442.2969488} introduces a convolutional neural network that operates directly on graphs of arbitrary sizes and shapes. A scalable approach \cite{DBLP:journals/corr/KipfW16} chooses the convolutional architecture via a localized approximation of spectral graph convolutions, which is an efficient variant and can operate on graphs directly as well. However, these methods can only be implemented on undirected graphs. Previously, in form of recurrent neural networks, Graph Neural Networks (GNNs) \cite{1555942,4700287} are proposed to operate on directed graphs. As a modification of GNN, gated GNN \cite{DBLP:journals/corr/LiTBZ15} uses gated recurrent units and employs back-propagation through time (BPTT) to compute gradients. Recently, GNN is broadly applied for the different tasks, e.g., script event prediction \cite{EEG2018}, situation recognition \cite{8237710}, and image classification \cite{8099493}.

\section{The Proposed Method}

In this section, we introduce the proposed SR-GNN which applies graph neural networks into session-based recommendation. We formulate the problem at first, then explain how to construct the graph from sessions, and finally describe the SR-GNN method thoroughly.

\subsection{Notations}

Session-based recommendation aims to predict which item a user will click next, solely based on the user's current sequential session data without accessing to the long-term preference profile. Here we give a formulation of this problem as below.

In session-based recommendation, let $V = \{v_1, v_2,\dots,v_m\}$ denote the set consisting of all unique items involved in all the sessions. An anonymous session sequence $s$ can be represented by a list $s$ $=$ $[v_{s,1}, v_{s,2},\dots, v_{s,n}]$ ordered by timestamps, where $v_{s,i} \in V$ represents a clicked item of the user within the session $s$. The goal of the session-based recommendation is to predict the next click, i.e. the sequence label, $v_{s,n+1}$ for the session $s$. Under a session-based recommendation model, for the session $s$, we output probabilities $\hat{\mathbf{y}}$ for all possible items, where an element value of vector $\hat{\mathbf{y}}$ is the recommendation score of the corresponding item. The items with top-$K$ values in $\hat{\mathbf{y}}$ will be the candidate items for recommendation.

\subsection{Constructing Session Graphs}

Each session sequence $s$ can be modeled as a directed graph $\mathcal{G}_s = (\mathcal{V}_s, \mathcal{E}_s)$. In this session graph, each node represents an item $v_{s,i} \in V$. Each edge $(v_{s,i-1}, v_{s,i}) \in \mathcal{E}_s$ means that a user clicks item $v_{s,i}$ after $v_{s,i-1}$ in the session $s$. Since several items may appear in the sequence repeatedly, we assign each edge with a normalized weighted, which is calculated as the occurrence of the edge divided by the outdegree of that edge's start node. We embed every item $v \in V$ into an unified embedding space and the node vector $\mathbf{v} \in \mathbb{R}^d$ indicates the latent vector of item $v$ learned via graph neural networks, where $d$ is the dimensionality. Based on node vectors, each session $s$ can be represented by an embedding vector $\mathbf{s}$, which is composed of node vectors used in that graph.

\subsection{Learning Item Embeddings on Session Graphs}

Then, we present how to obtain latent vectors of nodes via graph neural networks. The vanilla graph neural network is proposed by \citeauthor{4700287} (\citeyear{4700287}), extending neural network methods for processing the graph-structured data. \citeauthor{DBLP:journals/corr/LiTBZ15} (\citeyear{DBLP:journals/corr/LiTBZ15}) further introduce gated recurrent units and propose gated GNN. Graph neural networks are well-suited for session-based recommendation, because it can automatically extract features of session graphs with considerations of rich node connections. We first demonstrate the learning process of node vectors in a session graph. Formally, for the node $v_{s,i}$ of graph $\mathcal{G}_s$, the update functions are given as follows:
\begin{align}
	\mathbf{a}^t_{s,i} & = \mathbf{A}_{s,i:} \left[\mathbf{v}^{t-1}_1, \dots ,\mathbf{v}^{t-1}_n\right]^\top \mathbf{H} + \mathbf{b}, \label{eq:node-representation}\\
	\mathbf{z}^t_{s,i} & = \sigma\left(\mathbf{W}_z\mathbf{a}^t_{s,i}+\mathbf{U}_z\mathbf{v}^{t-1}_{i}\right), \label{eq:update-gate}\\
	\mathbf{r}^t_{s,i} & = \sigma\left(\mathbf{W}_r\mathbf{a}^t_{s,i}+\mathbf{U}_r\mathbf{v}^{t-1}_{i}\right), \label{eq:reset-gate}\\
	\widetilde{\mathbf{v}^t_{i}} & = \tanh\left(\mathbf{W}_o \mathbf{a}^t_{s,i}+\mathbf{U}_o \left(\mathbf{r}^t_{s,i} \odot \mathbf{v}^{t-1}_{i}\right)\right), \label{eq:candidate-state}\\
	\mathbf{v}^t_{i} & = \left(1-\mathbf{z}^t_{s,i} \right) \odot \mathbf{v}^{t-1}_{i} + \mathbf{z}^t_{s,i} \odot \widetilde{\mathbf{v}^t_{i}} , \label{eq:final-state}
\end{align}
where $\mathbf{H}\in \mathbb{R}^{d \times 2d}$ controls the weight, $\mathbf{z}_{s,i}$ and $\mathbf{r}_{s,i}$ are the reset and update gates respectively, $\left[\mathbf{v}^{t-1}_1, \dots , \mathbf{v}^{t-1}_n\right]$ is the list of node vectors in session $s$, $\sigma(\cdot)$ is the sigmoid function, and $\odot$ is the element-wise multiplication operator. $\mathbf{v}_{i} \in \mathbb{R}^d$ represents the latent vector of node $v_{s,i}$. The connection matrix $\mathbf{A}_s \in \mathbb{R}^{n \times 2n}$ determines how nodes in the graph communicate with each other and $\mathbf{A}_{s,i:} \in \mathbb{R}^{1 \times 2n}$ are the two columns of blocks in $\mathbf{A}_s$ corresponding to node $v_{s,i}$.

\begin{figure}
	\centering
	\includegraphics[width=0.6\columnwidth]{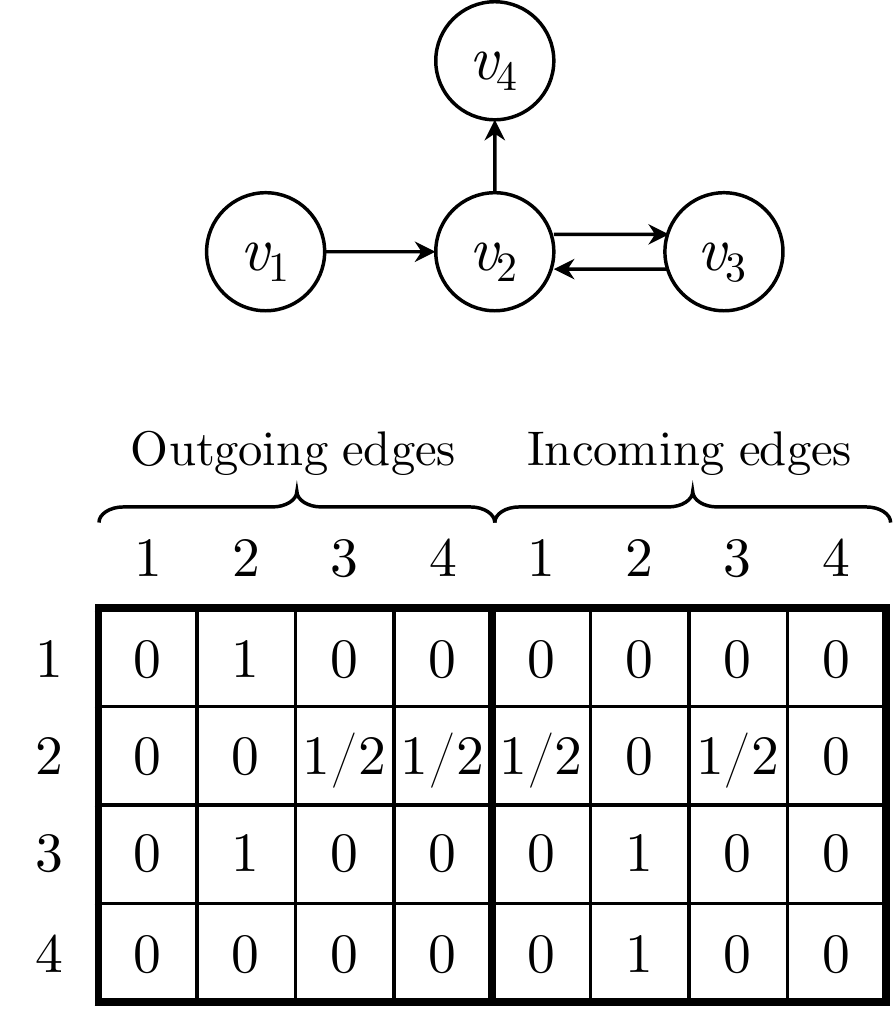}
	\caption{A example of a session graph and the connection matrix $\mathbf{A}_s$}
	\label{fig:connection-matrix}
\end{figure}

Here $\mathbf{A_s}$ is defined as the concatenation of two adjacency matrices $\mathbf{A}_s^{\text{(out)}}$ and $\mathbf{A}_s^{\text{(in)}}$, which represents weighted connections of outgoing and incoming edges in the session graph respectively. For example, consider a session $s=[v_1, v_2, v_3, v_2, v_4]$, the corresponding graph $\mathcal{G}_{s}$ and the matrix $\mathbf{A}_s$ are shown in Figure \ref{fig:connection-matrix}. Please note that SR-GNN can support different connection matrices $\mathbf{A}$ for various kinds of constructed session graphs. If different strategies of constructing the session graph are used, the connection matrix $\mathbf{A}_s$ will be changed accordingly. Moreover, when there exists content features of node, such as descriptions and categorical information, the method can be further generalized. To be specific, we can concatenate features with node vector to deal with such information.

For each session graph $\mathcal{G}_s$, the gated graph neural network proceeds nodes at the same time. Eq. (\ref{eq:node-representation}) is used for information propagation between different nodes, under restrictions given by the matrix $\mathbf{A}_s$. Specifically, it extracts the latent vectors of neighborhoods and feeds them as input into the graph neural network. Then, two gates, i.e. update and reset gate, decide what information to be preserved and discarded respectively. After that, we constructs the candidate state by the previous state, the current state, and the reset gate as described in Eq. (\ref{eq:candidate-state}). The final state is then the combination of the previous hidden state and the candidate state, under the control of the update gate. After updating all nodes in session graphs until convergence, we can obtain the final node vectors.

\subsection{Generating Session Embeddings}

Previous session-based recommendation methods always assume there exists a distinct latent representation of user for each session. On the contrary, the proposed SR-GNN method does not make any assumptions on that vector. Instead, a session is represented directly by nodes involved in that session. To better predict the users' next clicks, we plan to develop a strategy to combine long-term preference and current interests of the session, and use this combined embedding as the session embedding.

After feeding all session graphs into the gated graph neural networks, we obtain the vectors of all nodes. Then, to represent each session as an embedding vector $\mathbf{s} \in \mathbb{R}^d$, we first consider the local embedding $\mathbf{s}_\text{l}$ of session $s$. For session $s=[v_{s,1}, v_{s,2}, \dots, v_{s,n}]$, the local embedding can be simply defined as $\mathbf{v}_{n}$ of the last-clicked item $v_{s,n}$, i.e. $\mathbf{s}_\text{l} = \mathbf{v}_{n}$.

Then, we consider the global embedding $\mathbf{s}_\text{g}$ of the session graph $\mathcal{G}_s$ by aggregating all node vectors. Consider information in these embedding may have different levels of priority, we further adopt the soft-attention mechanism to better represent the global session preference:
\begin{equation}
	\begin{aligned}
		\alpha_i &= \mathbf{q}^\top \, \sigma(\mathbf{W}_1 \mathbf{v}_{n} + \mathbf{W}_2 \mathbf{v}_{i} + \mathbf{c}), \\
		\mathbf{s}_\text{g} & = \sum\limits_{i = 1}^{n} {\alpha_i \mathbf{v}_{i}},
	\end{aligned}
\end{equation}
where parameters $\mathbf{q} \in \mathbb{R}^{d}$ and $\mathbf{W}_1, \mathbf{W}_2 \in \mathbb{R}^{d \times d}$ control the weights of item embedding vectors.

Finally, we compute the hybrid embedding $\mathbf{s}_\text{h}$ by taking linear transformation over the concatenation of the local and global embedding vectors:
\begin{equation}
	\mathbf{s}_\text{h} = \mathbf{W}_3 \left[\mathbf{s}_\text{l}; \mathbf{s}_\text{g}\right],
\end{equation}
where matrix $\mathbf{W}_3 \in \mathbb{R}^{d \times 2d}$ compresses two combined embedding vectors into the latent space $\mathbb{R}^d$.

\subsection{Making Recommendation and Model Training}
After obtained the embedding of each session, we compute the score $\hat{\mathbf{z}_i}$ for each candidate item $v_i \in V$ by multiplying its embedding $\mathbf{v}_i$ by session representation $\mathbf{s}_\text{h}$, which can be defined as:
\begin{equation}
	\hat{\mathbf{z}_i} = \mathbf{s}_\text{h}^\top \, \mathbf{v}_i.
\end{equation}

Then we apply a softmax function to get the output vector of the model $\hat{\mathbf{y}}$:
\begin{equation}
	\hat{\mathbf{y}} = \operatorname{softmax}\left( \hat{\mathbf{z}} \right),
\end{equation}
where $\hat{\mathbf{z}} \in \mathbb{R}^m$ denotes the recommendation scores over all candidate items and $\hat{\mathbf{y}} \in \mathbb{R}^m$ denotes the probabilities of nodes appearing to be the next click in session $s$.

For each session graph, the loss function is defined as the cross-entropy of the prediction and the ground truth. It can be written as follows:
\begin{equation}
\mathcal{L}(\hat{\mathbf{y}}) = -\sum_{i = 1}^{m} \mathbf{y}_i \log{(\hat{\mathbf{y}_i})} + (1 - \mathbf{y}_i) \log{(1 - \hat{\mathbf{y}_i})},
\end{equation}
where $\mathbf{y}$ denotes the one-hot encoding vector of the ground truth item.

Finally, we use the Back-Propagation Through Time (BPTT) algorithm to train the proposed SR-GNN model. Note that in session-based recommendation scenarios, most sessions are of relatively short lengths. Therefore, it is suggested to choose a relatively small number of training steps to prevent overfitting.

\section{Experiments and Analysis}

In this section, we first describe the datasets, compared methods, and evaluation metrics used in the experiments. Then, we compare the proposed SR-GNN with other comparative methods. Finally, we make detailed analysis of SR-GNN under different experimental settings.

\subsection{Datasets}

We evaluate the proposed method on two real-world representative datasets, i.e. {\it Yoochoose}\footnote{\url{http://2015.recsyschallenge.com/challege.html}} and {\it Diginetica}\footnote{\url{http://cikm2016.cs.iupui.edu/cikm-cup}}. The Yoochoose dataset is obtained from the RecSys Challenge 2015, which contains a stream of user clicks on an e-commerce website within 6 months. The Diginetica dataset comes from CIKM Cup 2016, where only its transactional data is used.

\begin{table}
	\centering
	\caption{Statistics of datasets used in the experiments}
	\resizebox{\columnwidth}{!}{
	\begin{tabular}{cccc}
    	\toprule
		Statistics & {\it Yoochoose 1/64} & {\it Yoochoose 1/4} & {\it Diginetica} \\ \midrule
		\# of clicks & 557,248 & 8,326,407 & 982,961 \\
		\# of training sessions & 369,859 & 5,917,745 & 719,470 \\
		\# of test sessions & 55,898 & 55,898 & 60,858 \\
		\# of items & 16,766 & 29,618 & 43,097 \\
		Average length & 6.16  & 5.71  & 5.12 \\
		\bottomrule
	\end{tabular}
	}
	\label{tab:dataset-statistics}
\end{table}

For fair comparison, following \cite{Li:2017:NAS:3132847.3132926,Liu:2018:SSA:3219819.3219950}, we filter out all sessions of length 1 and items appearing less than 5 times in both datasets.
The remaining 7,981,580 sessions and 37,483 items constitute the Yoochoose dataset, while 204,771 sessions and 43097 items construct the Diginetica dataset. Furthermore, similar to \cite{Tan:2016:IRN:2988450.2988452}, we generate sequences and corresponding labels by splitting the input sequence. To be specific, we set the sessions of subsequent days as the test set for Yoochoose, and the sessions of subsequent weeks as the test set for Diginetiva. For example, for an input session $s = [v_{s,1}, v_{s,2}, \dots, v_{s,n}]$, we generate a series of sequences and labels $([v_{s,1}], v_{s,2}), ([v_{s,1}, v_{s,2}], v_{s,3}), \dots,$ $([v_{s,1}, v_{s,2}, \dots, v_{s,n-1}], v_{s,n})$, where $[v_{s,1}, v_{s,2}, \dots, v_{s,n-1}]$ is the generated sequence and $v_{s,n}$ denotes the next-clicked item, i.e. the label of the sequence. Following \cite{Li:2017:NAS:3132847.3132926,Liu:2018:SSA:3219819.3219950}, we also use the most recent fractions 1/64 and 1/4 of the training sequences of Yoochoose. The statistics of datasets are summarized in Table \ref{tab:dataset-statistics}.

\subsection{Baseline Algorithms}

To evaluate the performance of the proposed method, we compare it with the following representative baselines:

\begin{itemize}
	\item {\bf POP} and {\bf S-POP} recommend the top-$N$ frequent items in the training set and in the current session respectively.
	\item {\bf Item-KNN} \cite{Sarwar2001} recommends items similar to the previously clicked item in the session, where similarity is defined as the cosine similarity between the vector of sessions.
	\item {\bf BPR-MF} \cite{rendle2009bpr} optimizes a pairwise ranking objective function via stochastic gradient descent.
	\item {\bf FPMC} \cite{rendle2010factorizing} is a sequential prediction method based on markov chain.
	\item {\bf GRU4REC} \cite{DBLP:journals/corr/HidasiKBT15} uses RNNs to model user sequences for the session-based recommendation.
	\item {\bf NARM} \cite{Li:2017:NAS:3132847.3132926} employs RNNs with attention mechanism to capture the user's main purpose and sequential behavior.
	\item {\bf STAMP} \cite{Liu:2018:SSA:3219819.3219950} captures users' general interests of the current session and current interests of the last click.
\end{itemize}

\subsection{Evaluation Metrics}

Following metrics are used to evaluate compared methods.

{\bf P@20} (Precision) is widely used as a measure of predictive accuracy. It represents the proportion of correctly recommended items amongst the top-$20$ items.

{\bf MRR@20} (Mean Reciprocal Rank) is the average of reciprocal ranks of the correctly-recommended items. The reciprocal rank is set to 0 when the rank exceeds 20. The MRR measure considers the order of recommendation ranking, where large MRR value indicates that correct recommendations in the top of the ranking list.

\subsection{Parameter Setup}

Following previous methods \cite{Li:2017:NAS:3132847.3132926,Liu:2018:SSA:3219819.3219950}, we set the dimensionality of latent vectors $d=100$ for both datasets. Besides, we select other hyper-parameters on a validation set which is a random $10\%$ subset of the training set. All parameters are initialized using a Gaussian distribution with a mean of 0 and a standard deviation of 0.1. The mini-batch Adam optimizer is exerted to optimize these parameters, where the initial learning rate is set to 0.001 and will decay by 0.1 after every 3 epochs. Moreover, the batch size and the L2 penalty is set to 100 and $10^{-5}$ respectively.

\subsection{Comparison with Baseline Methods}

To demonstrate the overall performance of the proposed model, we compare it with other state-of-art session-based recommendation methods. The overall performance in terms of {\sc P@20} and {\sc MRR@20} is shown in Table \ref{tab:result-baseline-algorithms}, with the best results highlighted in boldface. Please note that, as in \cite{Li:2017:NAS:3132847.3132926}, due to insufficient memory to initialize FPMC, the performance on Yoochoose 1/4 is not reported.

\begin{table}
	\centering
	\caption{The performance of SR-GNN with other baseline methods over three datasets}
	\resizebox{\columnwidth}{!}{
	\begin{tabular}{ccccccc}
	\toprule
	\multirow{2}[0]{*}{Method} & \multicolumn{2}{c}{Yoochoose 1/64} & \multicolumn{2}{c}{Yoochoose 1/4} & \multicolumn{2}{c}{Diginetica} \\ \cmidrule(lr){2-3} \cmidrule(lr){4-5} \cmidrule(lr){6-7}
		& P@20 & MRR@20 & P@20 & MRR@20 & P@20 & MRR@20 \\ \midrule
	POP & 6.71  & 1.65  & 1.33  & 0.30  & 0.89  & 0.20 \\
    S-POP & 30.44 & 18.35 & 27.08 & 17.75 & 21.06 & 13.68 \\
    Item-KNN & 51.60 & 21.81 & 52.31 & 21.70  & 35.75 & 11.57 \\
    BPR-MF & 31.31 & 12.08 & 3.40  & 1.57  & 5.24  & 1.98 \\
    FPMC & 45.62 & 15.01 & -- & -- & 26.53 & 6.95 \\
    GRU4REC & 60.64 & 22.89 & 59.53 & 22.60 & 29.45 & 8.33 \\
    NARM & 68.32 & 28.63 & 69.73 & 29.23 & 49.70 & 16.17 \\
    STAMP & 68.74 & 29.67 & 70.44 & 30.00  & 45.64 & 14.32 \\
    SR-GNN & {\bf 70.57} & {\bf 30.94} & {\bf 71.36} & {\bf 31.89} & {\bf 50.73} & {\bf 17.59} \\
	\bottomrule
	\end{tabular}
	}
	\label{tab:result-baseline-algorithms}
\end{table}

SR-GNN aggregates separated session sequences into graph-structured data. In this model, we jointly consider the global session preference as well as the local interests. According to the experiments, it is obvious that the proposed SR-GNN method achieves the best performance on all three datasets in terms of P@20 and MRR@20. This verifies the effectiveness of the proposed method.

Regarding those traditional algorithms like POP and S-POP, their performance is relatively poor. Such simple models make recommendations solely based on repetitive co-occurred items or successive items, which is problematic in session-based recommendation scenarios. Even so, S-POP still outperforms its opponents such as POP, BPR-MF, and FPMC, demonstrating the importance of session contextual information. Item-KNN achieves better results than FPMC which is based on Markov chains. Please note that, Item-KNN only utilizes the similarity between items without considering sequential information. This indicates that the assumption on the independence of successive items, which traditional MC-based methods mostly rely on, is not realistic.

Neural-network-based methods, such as NARM and STAMP, outperform the conventional methods, demonstrating the power of adopting deep learning in this domain. Short/long-term memory models, like GRU4REC and NARM, use recurrent units to capture a user's general interest while STAMP improves the short-term memory by utilizing the last-clicked item. Those methods explicitly model the users' global behavioral preferences and consider transitions between users' previous actions and the next click, leading to superior performance against these traditional methods. However, their performance is still inferior to that of the proposed method. Compared with the state-of-art methods like NARM and STAMP, SR-GNN further considers transitions between items in a session and thereby models every session as a graph, which can capture more complex and implicit connections between user clicks. Whereas in NARM and GRU4REC, they explicitly model each user and obtain the user representations through separated session sequences, with possible interactive relationships between items ignored. Therefore, the proposed model is more powerful to model session behavior.

Besides, SR-GNN adopts the soft-attention mechanism to generate a session representation which can automatically select the most significant item transitions, and neglect noisy and ineffective user actions in the current session. On the contrary, STAMP only uses the transition between the last-clicked item and previous actions, which may not be sufficient. Other RNN models, such as GRU4REC and NARM, fail to select impactful information during the propagation process as well. They use all previous items to obtain a vector representing the user's general interest. When a user's behavior is aimless, or his interests drift quickly in the current session, conventional models are ineffective to cope with noisy sessions.

\subsection{Comparison with Variants of Connection Schemes}

The proposed SR-GNN method is flexible in constructing connecting relationships between items in the graph. Since user behavior in sessions is limited, we propose in this section another two connection variants in order to augment limited relationships between items in each session graph. Firstly, we aggregate all session sequences together and model them as a directed whole item graph, which is termed as the global graph hereafter. In the global graph, each node denotes a unique item, and each edge denotes a directed transition from one item to another. Secondly, we model all high-order relationships between items within one session as direct connections explicitly. In summary, the following two connection schemes are proposed to compare with SR-GNN:
\begin{itemize}
	\item SR-GNN with normalized global connections (SR-GNN-NGC) replaces the connection matrix with edge weights extracted from the global graph on the basis of SR-GNN.
	\item SR-GNN with full connections (SR-GNN-FC) represents all higher-order relationships using boolean weights and appends its corresponding connection matrix to that of SR-GNN.
\end{itemize}

\begin{figure}
	\centering
	\subfloat[P@20]{
		\includegraphics[width=0.47\columnwidth]{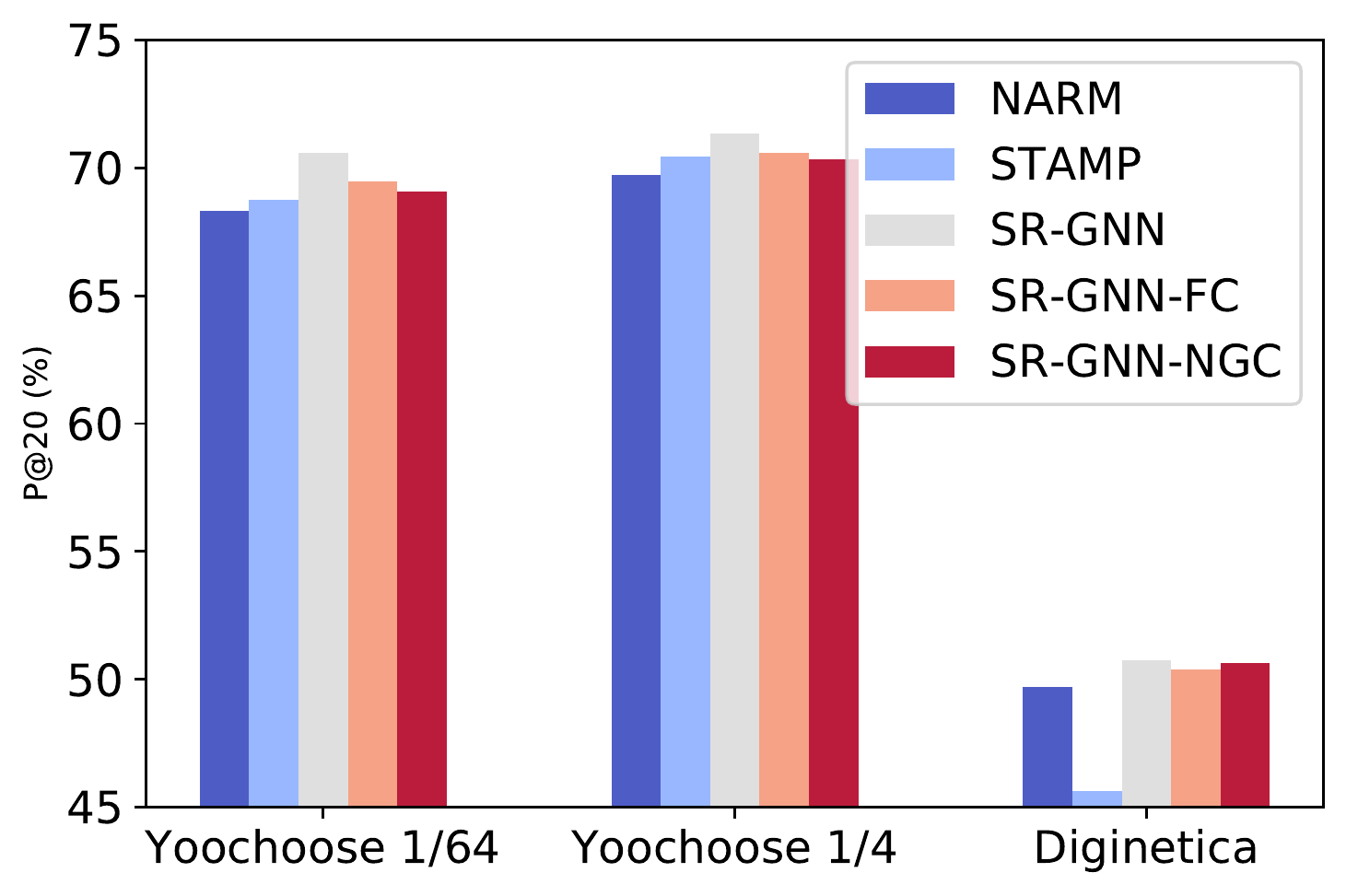}
	}\hfill
	\subfloat[MRR@20]{
		\includegraphics[width=0.47\columnwidth]{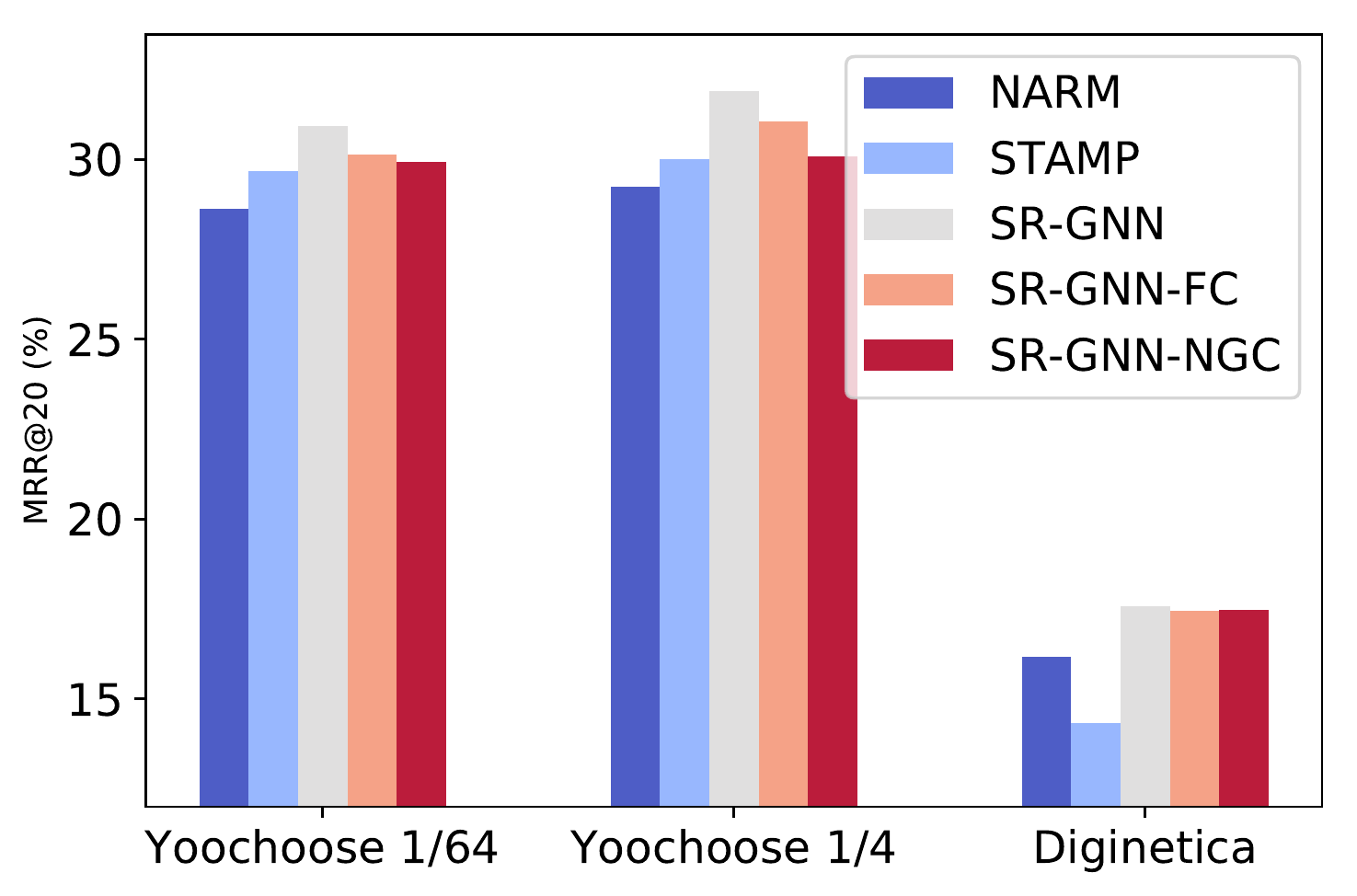}
	}\\
	\caption{The performance of different connection schemes}
	\label{fig:different-connection-schemes}
\end{figure}

The results of different connection schemes are shown in Figure \ref{fig:different-connection-schemes}. From the figures, it is seen that all three connection schemes achieve better or almost the same performance as the state-of-the-art STAMP and NARM methods, confirming the usefulness of modeling sessions as graphs.

Compared with SR-GNN, for each session, SR-GNN-NGC takes the impact of other sessions into considerations in addition to items in the current session, which subsequently reduces the influence of edges that are connected to nodes with high degree within the current session graph. Such a fusion method notably affects the integrity of the current session, especially when the weight of the edge in the graph varies, leading to performance downgrade.

In regard to SR-GNN and SR-GNN-FC, the former one only models the exact relationship between consecutive items, and the latter one further explicitly regards all high-order relationships as direct connections. It is reported that SR-GNN-FC performs worse than SR-GNN, though the experimental results of the two methods are not of much difference. Such a small difference in results suggests that in most recommendation scenarios, not every high-order transitions can be directly converted to straight connections and intermediate stages between high-order items are still necessities. For instance, considering that the user has viewed the following pages when browsing a website: $A \rightarrow B \rightarrow C$, it is not appropriate to recommend page $C$ directly after $A$ without intermediate page $B$, due to the lack of a direct connection between $A$ and $C$.

\subsection{Comparison with Different Session Embeddings}

\begin{figure}
	\centering
	\subfloat[P@20]{
		\includegraphics[width=0.47\columnwidth]{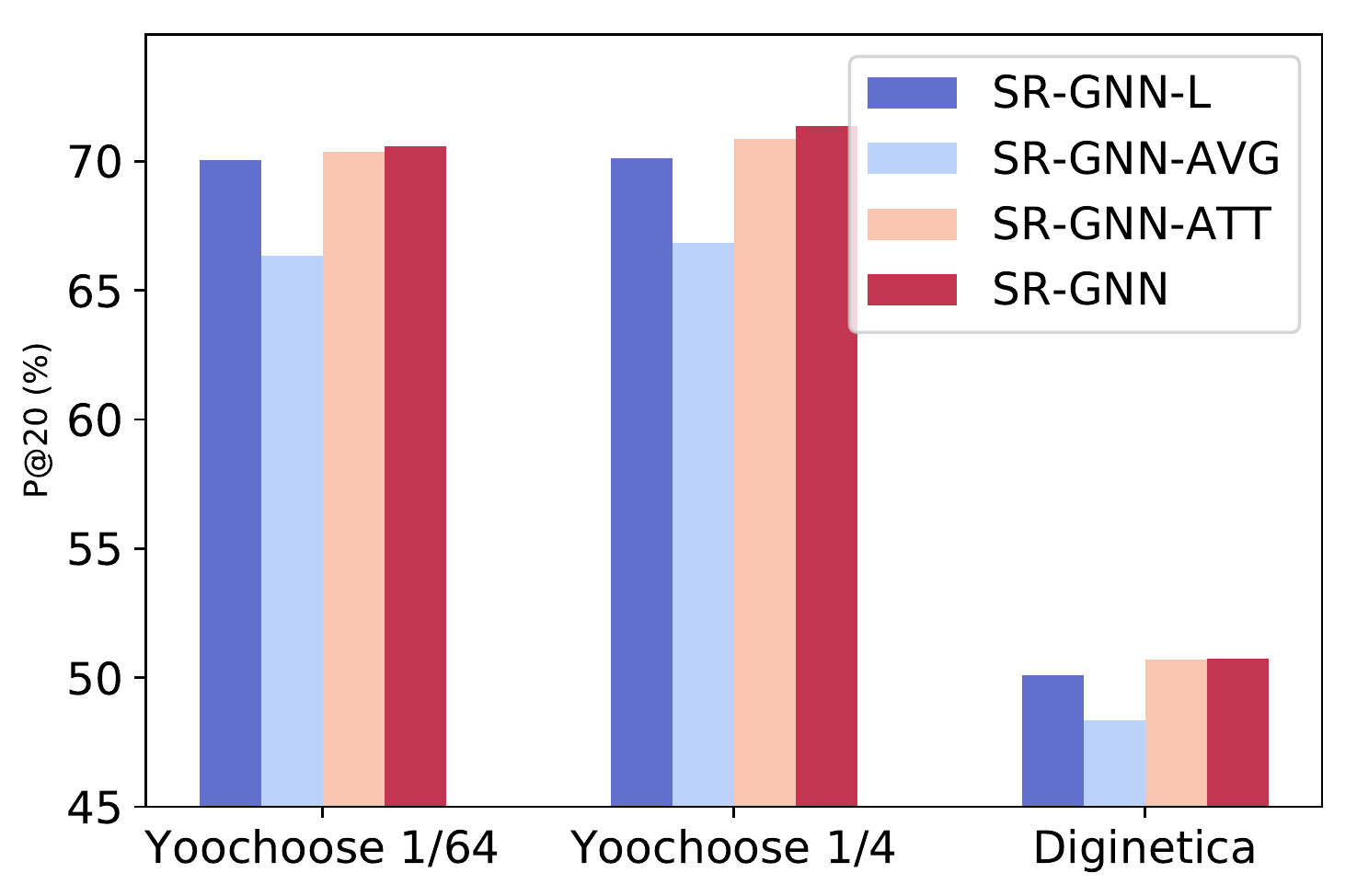}
	}\hfill
	\subfloat[MRR@20]{
		\includegraphics[width=0.47\columnwidth]{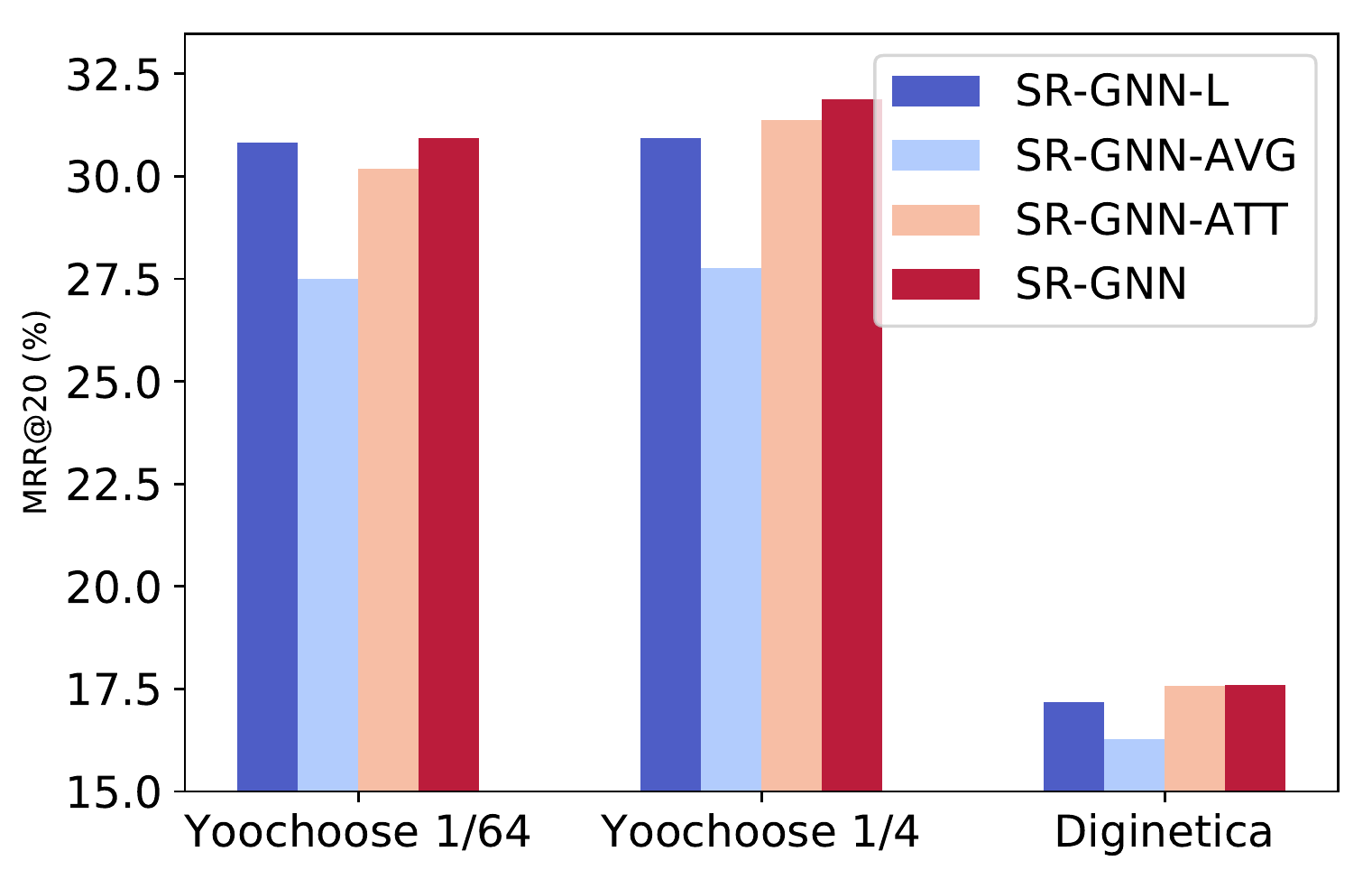}
	}\\
	\caption{The performance of different session representations}
	\label{fig:different-session-representations}
\end{figure}

We compare the session embedding strategy with the following three approaches: (1) local embedding only (SR-GNN-L), (2) global embedding with average pooling (SR-GNN-AVG), and (3) global embedding with the attention mechanism (SR-GNN-ATT). The results of methods with three different embedding strategies are given in Figure \ref{fig:different-session-representations}.

From the figures, it can be observed that the hybrid embedding method SR-GNN achieves best results on all three datasets, which validates the importance of explicitly incorporating current session interests with the long-term preference. Furthermore, the figures show that SR-GNN-ATT performs better than SR-GNN-AVG with average pooling on three datasets. It indicates that the session may contain some noisy behavior, which cannot be treated independently. Besides, it is shown that attention mechanisms are helpful in extracting the significant behavior from the session data to construct the long-term preference.

Please note that SR-GNN-L, a downgraded version of SR-GNN, still outperforms SR-GNN-AVG and achieves almost the same performance as that of SR-GNN-ATT, supporting that both the current interest and long-term preference are crucial for session-based recommendation.

\subsection{Analysis on Session Sequence Lengths}

We further analyze the capability of different models to cope with sessions of different lengths. For comparison, we partition sessions of Yoochoose 1/64 and Diginetica into two groups, where ``Short'' indicates that the length of sessions is less than or equal to 5, while each session has more than 5 items in ``Long''. The pivot value 5 is chosen because it is the closest integer to the average length of total sessions in all datasets. The percentages of session belonging to short group and long group are 0.701 and 0.299 on the Yoochoose data, and 0.764 and 0.236 on the Diginetica data. For each method, we report the results evaluated in terms of P@20 in Table \ref{tab:result-different-session-lengths}.

Our proposed SR-GNN and its variants perform stably on two datasets with different session lengths. It demonstrates the superior performance of the proposed method and the adaptability of  graph neural networks in session-based recommendation. On the contrary, the performance of STAMP changes greatly in short and long groups. STAMP \cite{Liu:2018:SSA:3219819.3219950} explains such a difference according to replicated actions. It adopts the attention mechanism, so replicated items can be ignored when obtaining user representations. Similar to STAMP, on Yoochoose, NARM achieves good performance on the short group, but the performance drops quickly with the length of the sessions increasing, which is partially because RNN models have difficulty in coping with long sequences.

Then we analyze the performance of SR-GNN-L, SR-GNN-ATT, and SR-GNN with different session representations. These three methods achieve promising results comparing with STAMP and NARM. It is probably because that based on the learning framework of graph neural networks, our methods can attain more accurate node vectors. Such node embedding not only captures the latent features of nodes but also models the node connections globally. On such basis, the performance is stable among variants of SR-GNN, while the performance of two state-of-art methods fluctuate considerably on short and long datasets. Moreover, the table shows that SR-GNN-L can also achieve good results, although this variant only uses local session embedding vectors. It is maybe because that SR-GNN-L also implicitly considers the properties of the first-order and higher-order nodes in session graphs. Such results are also validated by Figure \ref{fig:different-session-representations}, where both SR-GNN-L and SR-GNN-ATT achieve the close-to-optimal performance.

\begin{table}
	\centering
	\caption{The performance of different methods with different session lengths evaluated in terms of P@20}
	\resizebox{0.75\columnwidth}{!}{
    \begin{tabular}{ccccc}
		\toprule
    	\multirow{2}[0]{*}{Method} & \multicolumn{2}{c}{Yoochoose 1/64} & \multicolumn{2}{c}{Diginetica} \\ \cmidrule(lr){2-3} \cmidrule(lr){4-5}
    		& Short & Long & Short & Long \\ \midrule
		NARM  & {\bf 71.44} & 60.79 & {\bf 51.22} & 45.75 \\
		STAMP & 70.69 & 64.73 & 47.26 & 40.39 \\ \midrule
		SR-GNN-L & 70.11 & 69.73 & 49.04 & 50.97 \\
		SR-GNN-ATT & 70.31 & 70.64 & 50.35 & 51.05 \\
		SR-GNN & 70.47 & {\bf 70.70} & 50.49 & {\bf 51.27} \\
		\bottomrule
		\end{tabular}
	}
	\label{tab:result-different-session-lengths}
\end{table}

\section{Conclusions}

Session-based recommendation is indispensable where users' preference and historical records are hard to obtain. This paper presents a novel architecture for session-based recommendation that incorporates graph models into representing session sequences. The proposed method not only considers the complex structure and transitions between items of session sequences, but also develops a strategy to combine long-term preferences and current interests of sessions to better predict users' next actions. Comprehensive experiments confirm that the proposed algorithm can consistently outperform other state-of-art methods.

\section*{Acknowledgements}

The first two authors Shu Wu and Yuyuan Tang contribute to this work equally. The work is done during the internship of Yanqiao Zhu and Yuyuan Tang at CRIPAC, CASIA. The correspondence author is Yanqiao Zhu.

This work is jointly supported by National Natural Science Foundation of China (61772528), National Key Research and Development Program (2016YFB1001000), and National Natural Science Foundation of China (U1435221).

\small
\bibliography{aaai}
\bibliographystyle{aaai}

\end{document}